\documentclass[a4paper,11pt]{article}
\usepackage{jheppub} 

\usepackage{dirtytalk}
\usepackage{orcidlink}
\usepackage{diagbox}

\newcommand{\Z}{{\mathbb{Z}}}

\newcommand{\U}{\mathrm{U}}

\newcommand{\pqty}[1]{\left( #1 \right)}
\newcommand{\bqty}[1]{\left[ #1 \right]}
\newcommand{\abs}[1]{\left\lvert #1\right\rvert}

\newcommand{\expval}[1]{\langle #1 \rangle}

\newcommand {\dv}[3][ ]{
  \ifx #1 { }
    \frac{d #2}{d #3}
  \else
    \frac{d^{#1} #2}{d #3^{#1}}
  \fi
}
\newcommand {\pdv}[3][ ]{
  \ifx #1 { }
    \frac{\partial #2}{\partial #3}
  \else
    \frac{\partial^{#1} #2}{\partial #3^{#1}}
  \fi
}
\newcommand {\fdv}[3][ ]{
  \ifx #1 { }
    \frac{\delta #2}{\delta #3}
  \else
    \frac{\delta^{#1} #2}{\delta #3^{#1}}
  \fi
}

\title{\boldmath The finite temperature ground state energy of the confining string in three-dimensional \texorpdfstring{$U(1)$}{U(1)} gauge theory}

\author{M.~Caselle,\orcidlink{0000-0001-5488-142X}}
\author{A.~Mariani.\orcidlink{0000-0002-8830-920X}}

\affiliation{Physics Department, University of Turin \& INFN, Turin unit,\\ Via Pietro Giuria 1,\\ I-10125 Turin, Italy}

\emailAdd{caselle@to.infn.it}
\emailAdd{a.mariani@unito.it}

\abstract{The three-dimensional $\U(1)$ gauge theory displays a peculiar form of confinement which does not fit in the standard paradigm of effective string theory. In this work, we report results of numerical computations of the ground state energy of the confining string in the lattice theory at finite temperature, which we compared with various theoretical predictions. We find markedly different behavior compared to non-Abelian theories. For small values of the bare coupling, far from the continuum, the results are well-described by the Nambu-Goto string. In an intermediate region, all known theoretical predictions are incompatible with the numerical data. For large bare coupling, close to the continuum, we find remarkable agreement with a formula recently derived by Aharony, Barel and Sheaffer.}

\begin{document}
\maketitle
\flushbottom

\section{Introduction}
\label{sec:intro}

In the confining phase of a gauge theory, the energy flux between static charged particles forms a thin flux tube, also known as the \say{confining string}. To leading order, the potential energy $V(R)$ rises linearly with the separation $R$ between the charged particles. The potential also receives subleading corrections, most famously the universal $1/R$ \say{L\"uscher term} \cite{luscher1981, LSW1980}. 

Usually, the behavior of the confining string is described by an effective field theory known as \say{effective string theory} \cite{Nambu:1974zg,Goto:1971ce,luscher1981,LSW1980,Polchinski:1991ax,Aharony:2013ipa, Brandt:2016xsp, Caselle:2021eir}. In recent years, advances in effective string theory have allowed the calculation of several orders of subleading corrections to the interquark potential \cite{Meyer:2006qx,Luscher:2004ib, Aharony:2009gg, Aharony:2011gb, Dubovsky:2012sh, Billo:2012da, Gliozzi:2012cx, Aharony:2013ipa}. In this paradigm, the confining string is found to be described by small corrections around the Nambu-Goto string. Very recently, numerical simulations in the context of lattice field theory have quantitatively confirmed these predictions for the Ising gauge theory \cite{Baffigo:2023rin} as well as for non-Abelian gauge theories in three dimensions \cite{Caselle:2024zoh, Athenodorou:2011rx,Athenodorou:2016kpd, Caristo:2021tbk}.

The situation is rather different for $\U(1)$ gauge theory in three spacetime dimensions, which is the theory of interest in this work. This theory is remarkably well-understood theoretically \cite{Polyakov1977, GopfMack}. It is the only gauge theory for which confinement has been rigorously proven \cite{GopfMack}, and it is in fact confining for all values of the bare coupling. The nature of confinement in this theory is however rather different from the case of non-Abelian gauge theories. In particular, it is due to topological configurations, i.e. monopoles. A rather peculiar feature is that the string tension $\sigma$ and the mass gap $m$ do not scale commensurately in the continuum limit (like in non-Abelian gauge theories), but rather $m / \sqrt{\sigma} \to 0$ in the continuum limit. This poses a significant challenge for the effective string theory description of this model. In fact, a basic assumption of effective string theory (which is realized for non-Abelian gauge theories) is that, due to the mass gap, at sufficiently low energies the only relevant degrees of freedom are the massless Goldstone bosons of the broken transverse translations \cite{luscher1981, Dubovsky:2012sh}. However, for $\U(1)$ in three dimensions the mass gap becomes exponentially small and this assumption is no longer correct.

For these reasons, it is unclear to what extent the confining string in three-dimensional $\U(1)$ gauge theory may be described by effective string theory. In order to describe the confining string in this model, Polyakov \cite{Polyakov:1996nc} introduced the so-called \say{rigid string}. This was found to be in reasonable agreement with numerical data for the zero temperature string tension \cite{DifferentString}, but there are several issues with this description. We will describe this proposal and its difficulties in Section \ref{sec:theoretical predictions}. Thus we should also consider alternative descriptions of the confining string in this model. Of particular relevance for our results is a very recent analytical calculation of the interquark potential directly from \textit{microscopic} gauge theory variables \cite{Aharony:2024ctf}, which sidesteps the issues with effective string theory. This remarkable calculation is only possible due to our good theoretical understanding of three-dimensional $\U(1)$ gauge theory \cite{Polyakov1977, GopfMack}.

In this work, we compute the ground state energy of the confining string in three-dimensional $\U(1)$ gauge theory at finite temperature. As will be reviewed in Section \ref{sec:theoretical predictions}, the advantage of working at finite temperature is the suppression of boundary terms which would otherwise obscure the subleading bulk behavior of the string. In Section \ref{sec:threed u(1)} we discuss the three dimensional $\U(1)$ lattice gauge theory and set the conventions. In Section \ref{sec:theoretical predictions} we discuss the various theoretical predictions for the ground state energy. Then in Section \ref{sec:results} we explain the setup of our numerical simulations, as well as our results, which we will compare to the various theoretical predictions.

\section{The three-dimensional \texorpdfstring{$U(1)$}{U(1)} gauge theory}
\label{sec:threed u(1)}

In the standard formulation of $\U(1)$ lattice gauge theory, one assigns a $\U(1)$-valued variable $e^{i\theta_l}$ to each link $l$ of a cubic lattice. Choosing the Wilson action, the path integral is then given by
\begin{equation}
    Z = \int_0^{2\pi} \pqty{\prod_{l} d\theta_l} \exp{\bqty{\beta \sum_{p} \cos{\theta_p} }} .
\end{equation}
where $\theta_p$ is the standard plaquette variable, i.e. $\theta_p = \theta_1+\theta_2-\theta_3-\theta_4$ in terms of the four variables associated to the four links making up a plaquette. Here $\beta$ is a bare coupling. 

In three spacetime dimensions, the behaviour of the theory is well-understood theoretically \cite{Polyakov1977,GopfMack}. It is confining for all values of $\beta$, and it is the only gauge theory for which confinement has been rigorously proven \cite{GopfMack}. The continuum limit is obtained as $\beta \to \infty$. The scaling of the mass and string tension has also been predicted for large $\beta$:
\begin{align}
    \label{eq:mass prediction}
    a m &=  c_0 \sqrt{\beta} \exp{\bqty{-\pi^2 v_0 \beta}}  \ , \\
    \label{eq:string tension prediction}
    a^2 \sigma &= \frac{\widetilde{c}}{\pi \sqrt{2 \beta }} \exp{\bqty{-\pi^2 v_0\beta}} \ , 
\end{align}
where $a$ is the lattice spacing, $v_0 \approx 0.2527$ and $c_0,\widetilde{c}$ are constants. These results have found reasonable numerical confirmation~\cite{TepAthen, DifferentString, Loan2003, Loan:2003wy, Caselle:2025bgu}. Concretely, the best fit extrapolation to the numerical data for the Wilson action was computed in \cite{TepAthen} as
\begin{align}
    \label{eq:mass extrapolation}
    a m &= 59.4(1.1) \sqrt{\beta} \exp{\bqty{-2.633(10) \beta}}  \ , \\
    \label{eq:string tension extrapolation}
    a^2 \sigma &= \frac{11.5(6)}{\sqrt{\beta }} \exp{\bqty{-2.561(22)\beta}} \ .
\end{align}
We will make use of this extrapolation in Section \ref{sec:results}. The above scaling for the mass and string tension imply the following limits for the dimensionless ratios,
\begin{equation}
    \label{eq:ratio scaling}
    \frac{m}{\sqrt{\sigma}} \to 0  \ , \quad \quad \quad \frac{(a^2 \sigma) \beta}{(am)} \to \mathrm{constant} \neq 0 \ .
\end{equation}
Due to the above scaling relations, it is impossible to take a continuum limit where both the mass and string tension take finite, non-zero values (on the other hand, this \textit{can} be done in non-Abelian gauge theories). As a result, the theory admits several possible continuum limits depending on what variable is held constant \cite{TepAthen, MarianiThesis, Caselle:2025bgu}. We also mention that several works have studied various properties of this theory at finite temperature \cite{Coddington:1986jk, Chernodub1, Borisenko:2008sc, Borisenko:2010qe, Borisenko:2015jea, Caselle:2025bgu}.

In this work, we are interested in comparing numerical data with predictions for the energy of the confining string, so we consider the continuum limit where the string tension is held fixed. In this case, the mass goes to zero as $\beta\to \infty$. As we will see in Section \ref{sec:theoretical predictions}, this has important implications for the applicability of effective string theory to this model.

It should also be emphasized that the predictions for the mass, eq.\eqref{eq:mass prediction}, and the string tension, eq.\eqref{eq:string tension prediction}, are not on equal theoretical footing. While the prediction for the mass does not require further assumptions (it is in fact not difficult to compute higher order corrections \cite{MarianiThesis}), the prediction for the string tension rests on a semiclassical approximation and therefore will receive quantum corrections. Rigorous upper \cite{Ito:1981tq} and lower \cite{GopfMack} bounds have been derived to further reinforce the scaling eq.\eqref{eq:string tension prediction}, but they are not tight. For a thorough discussion of these points, the reader can consult Chapter 2 of \cite{MarianiThesis}. Very recently, for the first time, the leading-order quantum correction to the semiclassical string tension has been computed \cite{Aharony:2024ctf}. Their techniques also allow the calculation of the leading-order correction to the interquark potential, a result which we will discuss more thoroughly in Section \ref{sec:theoretical predictions}. 

Our analytical and numerical understanding of this theory rests on an exact duality transformation ~\cite{Banks:1977cc,Savit:1977fw,GopfMack}. The dual variables are integer-valued scalar fields $h_x \in \Z$ which live on the sites of the dual lattice. In terms of the dual variables, the partition function for the Wilson action is given by
\begin{equation}
    \label{eq:dual partition function}
    Z =\pqty{\prod_x \sum_{h_x \in \Z}} \prod_{\expval{xy}} I_{\abs{h_x-h_y}}(\beta) \ ,
\end{equation}
where the first product is over all (dual) lattice sites and then second product over all (dual) links $l$ connecting (dual) sites $x$ and $y$. The $I_n(\beta)$ are modified Bessel functions of the first kind. This is the theory which we will simulate numerically.

We will be interested in computing the ground state energy of the confining string at finite temperature. This means that we work on lattices with $N_t$ sites in the time direction and $L$ sites in each spatial direction, with $N_t \ll L$. The ground state energy of the string can be computed as follows. One first defines the Polyakov loop variable at each spatial site $\vec x$,
\begin{equation}
    P(\vec{x}) = \prod_{t=0}^{N_t-1} e^{i \theta_{t}(\vec{x},t)} \ .
\end{equation}
The correlator between two Polyakov loops can be expressed as a ratio of two partition functions \cite{Luscher:2004ib},
\begin{equation}
    \expval{P(\vec x) P(\vec x + R)} = \frac{Z(R)}{Z} \ ,
\end{equation}
where $Z$ is the usual partition function, i.e. eq.\eqref{eq:dual partition function}, and $Z(R)$ is the partition function with the insertion of two static charges separated by a distance $R$. The spectral expansion for the Polyakov loop correlator has been derived on rather general grounds \cite{Luscher:2004ib}. In three spacetime dimensions it reads
\begin{equation}
    \label{eq:luscher weisz correlator}
    \expval{ P(\vec{x})P^\dagger(\vec{x}+R)} = \frac{1}{\pi}  \sum_{n=0}^{\infty} \abs{v_n}^2 E_n K_{0}({E}_n R) \ ,
\end{equation}
where the $E_n= E_n(N_t)$ are the string energies, $\abs{v_n}^2$ are positive real numbers and $K_0$ is a modified Bessel function of the second kind. Note that $K_0(z) \approx \sqrt{\frac{\pi}{2z}} e^{-z}$ for large $z$. As described in Section \ref{sec:results}, we will use the prediction eq.\eqref{eq:luscher weisz correlator} to extract the ground state energy $E_0$.

Numerically, the ratio $Z(R)/Z$ can be efficiently computed using the so-called \say{snake algorithm} \cite{deForcrand:2000fi, deForcrand:2004jt}. Here we use a version where each replica is computed using hierarchical Metropolis updates \cite{CaselleHasenbusch}. For a thorough discussion of the application of the snake algorithm to this theory, see Chapter 2 of \cite{MarianiThesis}. Here we give only a brief outline. Since the ratio $Z(R)/Z$ is expected to have poor overlap, its direct calculation is likely difficult. To improve the overlap, one notes that the ratio can be factored as
\begin{equation}
    \label{eq:snake factorization}
    \frac{Z(R)}{Z} = \prod_{r = 0}^{R-1} \frac{Z(r+1)}{Z(r)} \ ,
\end{equation}
where each term $Z(r+1)/Z(r)$ has improved overlap. This ratio can again be factored in a similar manner to further improve the overlap. It is important to note that each term $Z(r+1)/Z(r)$ is computed independently, which greatly simplifies the error analysis. For this reason, assuming a similar number of Monte Carlo samples, the error on $Z(r+1)/Z(r)$ is expected to be roughly independent of $r$ (and therefore the error on $Z(R)/Z$ increases with $R$). For this reason, it is most natural to fit directly the ratios $Z(r+1)/Z(r)$. We will give further details on the fitting procedure in Section \ref{sec:results}.

\section{Theoretical predictions for the string ground state energy}
\label{sec:theoretical predictions}

In most cases, the behavior of the flux tube in the confining phase of a gauge theory is well-described by effective string theory \cite{Nambu:1974zg,Goto:1971ce,luscher1981,LSW1980,Polchinski:1991ax}. In this section, we will discuss only those aspects of effective string theory which are directly relevant for our work. For a more thorough discussion, we point the reader to the reviews \cite{Aharony:2013ipa, Brandt:2016xsp, Caselle:2021eir}.

In recent years, significant progress has been made in the understanding of effective string theory \cite{Meyer:2006qx,Luscher:2004ib, Aharony:2009gg, Aharony:2011gb, Dubovsky:2012sh, Billo:2012da, Gliozzi:2012cx, Aharony:2013ipa}. Special features of the string action, as well as the incorporation of Poincar\'e invariance of the underyling gauge theory allows an efficient organization of the derivative expansion of the effective theory. This means that, in three spacetime dimensions, the effective string action can be written as 
\begin{equation}
    \label{eq:effective string action}
    S_{EST}= \int d^2\xi \sqrt{g}\,\bqty{ \sigma +  \gamma_1\mathcal{R} + \gamma_2 K^2 +  \gamma_3 K^4  + \cdots } \ , 
\end{equation}
where $\sigma$ (the zero-temperature string tension) as well as the coefficients $\gamma_i$ are low-energy constants which are not predicted by the effective theory. Here $\xi$ are the two worldsheet coordinates (i.e. time and string arclength), $X_\mu(\xi)$ is the mapping of the world-sheet in target space, i.e. (after gauge-fixing) it is the field representing the Goldstone bosons of the broken transverse translation symmetry \cite{luscher1981, Dubovsky:2012sh}. The above terms are organized in a geometric fashion and can be constructed from the induced metric on the worldsheet,
\begin{equation}
    g_{\alpha\beta} = \partial_\alpha X_\mu \partial_\beta X^\mu \ .
\end{equation}
In these terms, one has $g \equiv \det g_{\alpha\beta}$ and then $\mathcal{R}$ is the Ricci scalar constructed from the metric, while $K$ is the extrinsic curvature.

To lowest order, the effective string action is given by the integral of $\sqrt{g} \equiv \sqrt{\det g_{\alpha\beta}}$, which corresponds simply to the Nambu-Goto action \cite{Nambu:1974zg, Goto:1971ce}. As it turns out, this action is exactly integrable and the Polyakov loop correlator at finite temperature can be computed exactly \cite{Dubovsky:2012sh, Caselle:2013dra, Luscher:2004ib, Billo:2005iv}. In $d$ spacetime dimensions, the ground state energy of the Nambu-Goto string is given by
\begin{equation}
    \label{eq:E0 Nambu-Goto}
    E_0(N_t) = \sigma N_t \sqrt{1-\frac{\pi (d-2)}{3 \sigma N_t^2}} \ .
\end{equation}
This formula can be used to estimate the location of the deconfinement transition (i.e. the point where the \say{finite-temperature string tension} $E_0(N_t)/N_t$ becomes zero), leading to the prediction \cite{Olesen:1985ej,Pisarski:1982cn}
\begin{equation}
    \label{eq:Nambu-Goto critical temperature}
    \frac{T_{c,NG}}{\sqrt{\sigma}}=\sqrt{\frac{3}{\pi(d-2)}} \ ,
\end{equation}
This is in rough quantitative agreement with numerical data for non-Abelian gauge theories as well as for the three-dimensional Ising model. However, eq.\eqref{eq:E0 Nambu-Goto} also predicts a second-order deconfinement transition with mean-field exponents, which is virtually always incorrect and in disagreement with the Svetitsky-Yaffe conjecture \cite{Svetitsky1982}. For the three-dimensional $\U(1)$ gauge theory, the prediction eq.\eqref{eq:Nambu-Goto critical temperature} is completely incorrect, since the actual critical temperature in units of the string tension $T_c/\sqrt{\sigma} \to \infty$ is infinite in the continuum limit \cite{Chernodub1, Borisenko:2008sc, Borisenko:2010qe, Borisenko:2015jea, Caselle:2025bgu}, i.e. the theory does not deconfine. This is already a hint that deviations from Nambu Goto are stronger in this theory compared to non-Abelian gauge theories.

As is clear from the expansion of the effective string action, eq.\eqref{eq:effective string action}, the Nambu-Goto string should be considered only the leading-order approximation to the actual effective string theory. However, the term proportional to $\gamma_1$ is a topological invariant and can be ignored since we do not expect topology-changing fluctuations. Moreover, the term proportional to $\gamma_2$ can be removed by a field redefinition \cite{Aharony:2013ipa} (although see the later discussion on this point). Therefore the lowest-order corrections to Nambu-Goto behavior start from the $\gamma_3$ term, which contributes to the energies at order $1/R^7$ (or $1/N_t^7$ in the finite temperature setting). This result is also known as \say{low energy universality} \cite{Dubovsky:2012sh}.

A significant problem in measuring these small corrections to the interquark potential is that they only correspond to the bulk behavior of the string. For a complete treatment of the problem, one should also include boundary terms. At zero temperature, these give contributions of the order $1/R^4$, thus overshadowing the corrections to Nambu-Goto \cite{Aharony:2010cx,Billo:2012da,Brandt:2010bw,Brandt:2017yzw,Brandt:2021kvt}. To overcome this difficulty, one may instead work at finite temperature, i.e. finite $N_t \ll L$ (which corresponds to the \say{closed string} channel), where boundary corrections can be neglected for $R \gg N_t$ \cite{Caselle:2011vk,Caselle:2021eir}. This is the choice that we have made in this work. 

Using S-matrix techniques it is possible to further constrain the low-energy constants \cite{Dubovsky:2012sh,EliasMiro:2019kyf}, thus greatly enhancing the predictive power of effective string theory. In particular, defining $\tilde\gamma_n{=}\gamma_n+(-1)^{(n+1)/2}\frac{1}{n 2^{3n-1}}$ one finds the inequalities
\begin{align}
    \label{eq:gamma_i inequalities}
    \tilde  \gamma_3&\geq 0 \ ,\nonumber \\ 
    \tilde \gamma_5&\geq 4 \tilde \gamma_3^2 - \frac{1}{64}\tilde \gamma_3 \ .
\end{align}
In particular, one must have $\gamma_3 \geq -\frac{1}{768}$. Moreover, the Thermodynamic Bethe Ansatz allows the calculation of the $\gamma_3$ and $\gamma_5$ corrections to the string ground state energy in the regime $R \gg N_t$ \cite{EliasMiro:2019kyf}:
\begin{equation}
    \label{eq:gamma3 gamma5 corrections}
    E_0(N_t) = \sigma N_t \sqrt{1-\frac{\pi}{3\sigma N_t^2}} -\frac{32 \pi^6 \gamma_3}{225 \sigma^3 N_t^7}-\frac{64 \pi^7 \gamma_3}{675 \sigma^4 N_t^9}-\frac{\frac{2 \pi^8 \gamma_3}{45} +\frac{32768 \pi^{10} \gamma_5}{3969}}{\sigma^5N_t^{11}} \ .
\end{equation}
Recently, this prediction has been found to be in agreement with numerical data for the Ising gauge theory \cite{Baffigo:2023rin} as well as for non-Abelian lattice gauge theories \cite{Caselle:2024zoh, Athenodorou:2011rx,Athenodorou:2016kpd, Caristo:2021tbk}, from which the $\gamma_3$ and $\gamma_5$ coefficients were estimated. In Section \ref{sec:results}, we will show a comparison of this prediction with numerical data for the three-dimensional $\U(1)$ gauge theory. 

The $\U(1)$ gauge theory in three dimensions however poses a set of non-trivial challenges to the above effective string theory picture.
In fact, the applicability of effective string theory rests on the assumption that the only relevant degrees of freedom below the confinement scale are the massless Goldstone bosons of the broken transverse translation symmetry due to the presence of the string \cite{luscher1981, Dubovsky:2012sh}. This assumption, however, breaks down for the three-dimensional $\U(1)$ gauge theory where, as we have seen in Section \ref{sec:threed u(1)}, one has an additional massless bulk degree of freedom in the continuum. On the other hand, at finite $\beta$ the mass is exponentially small but not zero, see eq.\eqref{eq:mass prediction}. Thus we expect that the predictions of standard effective string theory should be applicable at small $\beta$ where the mass is relatively large; as discussed in Section \ref{sec:results}, this is indeed what we find.

In the rest of this section, we will discuss two proposals to overcome this issue: one was proposed long ago by Polyakov, and is known as the \say{rigid string} \cite{Polyakov:1996nc}. The other was proposed very recently in \cite{Aharony:2024ctf}, and is based on the quantization of the modes on top of the semiclassical solution for the confining string which we discussed in Section \ref{sec:intro}.

\subsection{The rigid string prediction}

For the three-dimensional $\U(1)$ gauge theory, Polyakov \cite{Polyakov:1996nc} proposed that its string should be described as a \say{rigid string} (see also \cite{Polyakov:1986cs, Kleinert:1986bk, Alonso:1986gf, Antonov:1998kw}). This corresponds to the addition of an extrinsic curvature term $K^2$ to the Nambu-Goto action (like in eq.\eqref{eq:effective string action}). An appealing feature of this proposal is that $\gamma_2$, for dimensional reasons is expected to scale as $\gamma_2 \sim \sigma/m^2$ and therefore (from eq.\eqref{eq:ratio scaling}) the extrinsic curvature term is expected to dominate the effective string in the continuum. It can be shown  \cite{Braaten:1986bz, German:1989vk, Ambjorn:2014rwa, Nesterenko:1997ku, DifferentString}  that the contribution to the ground state energy of the rigidity term behaves as an additional massive boson, with mass $\widetilde{m}^2=\frac{\sigma}{2\gamma_2}$. The correction to the ground state energy due to this massive boson is easy to evaluate and in the closed string channel is given by:
\begin{equation}
    \label{eq:rigid string prediction}
    \Delta E_{0}(N_t) = -\frac{ \widetilde{m}}{\pi} \sum_{n=1}^\infty \frac{K_1(n \widetilde{m} N_t)}{n} \ ,
\end{equation}
where in this framework $\widetilde{m}$ should be considered as a free parameter which encodes the strength of the rigidity term. In the limit of large $\gamma_2$ the mass $\widetilde{m}$ vanishes and the tower of Bessel function can be resummed so as to give an additional \say{L\"uscher term} as expected for a free massless boson. When $\gamma_2$ is small $\widetilde{m}$ goes to infinity and the contribution is exponentially suppressed. It is useful to note that the rigid string correction comes with a negative sign exactly as the standard L\"uscher term of the Nambu-Goto string.

The \say{rigid string} prediction eq.\eqref{eq:rigid string prediction} was tested in \cite{DifferentString} by computing the interquark potential at zero temperature. It was found to be in reasonable agreement with the data, but the determination of $\widetilde{m}$ was affected by a large uncertainty; moreover, the fits were complicated by the presence of a boundary term which overshadows fine corrections to the interquark potential.
As already mentioned, in this work we have chosen a finite temperature setting in which the boundary term is suppressed with the precise goal of performing a more stringent test of this picture. As we will see in Section \ref{sec:results}, our data at finite temperature is incompatible with the rigid string prediction; in particular, we find very large values of $\widetilde{m}$ and accordingly $\gamma_2$ essentially vanishes, in agreement with low-energy universality.

\subsection{The prediction by Aharony, Barel, Sheaffer}

Very recently, a new calculation \cite{Aharony:2024ctf} has completely sidestepped the issues with effective string theory by computing the interquark potential in three-dimensional $\U(1)$ gauge theory directly from the microscopic variables. In other words, they computed the leading-order quantum correction to the semiclassical approximation of the string tension computed in \cite{Polyakov1977, Banks:1977cc, Muller1981, Ambjorn:1982ts, GopfMack} (see also the discussion in Section \ref{sec:threed u(1)}). In our notation, the ground state energy of the string given by this novel calculation \cite{Aharony:2024ctf} takes the form
\begin{equation}
    E_0(N_t) = \sigma N_t - \frac{\pi}{6 N_t} + \frac{1}{\pi N_t} \mathrm{Li}_2(e^{-m N_t}) \ ,
     \label{eq:aharony prediction}
\end{equation}
where $m$ is the mass gap and $\mathrm{Li}_2$ is the dilogarithm. For our range of values this is approximately an exponential, since for small values of $z$ one has $\mathrm{Li}_2(z) \approx z$. Note that in this expression there are no additional free parameters: both the mass and the string tension must coincide with their microscopic values approximately given by eq.s\eqref{eq:mass prediction}-\eqref{eq:string tension prediction}. This is thus a highly predictive expression. In Section \ref{sec:results}, we compare this prediction with numerical data.

It is interesting to compute the limiting cases of the dilogarithm. For large $m N_t$ (i.e. far from the continuum) one has $\mathrm{Li}_2(e^{-m N_t}) \to \mathrm{Li}_2(0) = 0$; in this case the correction vanishes and, as expected, we recover the standard linear plus L\"uscher term\footnote{Note however that, as discussed below, even if this result is consistent with our expectations, this region is far from the regime of applicability of eq.\eqref{eq:aharony prediction}.}. On the other hand, for small $m N_t$ (i.e. close to the continuum) the dilogarithm is quite relevant and in fact in the limit one has $\mathrm{Li}_2(e^{-m N_t}) \to \mathrm{Li}_2(1)=\frac{\pi^2}{6}$, thus, together with the prefactor, the dilogarithm term exactly cancels the L\"uscher term. 
A major consequence of this observation is that this model predicts the correct location of the deconfinement transition in the continuum limit. 
In fact looking at the zero of the \say{finite-temperature string tension} $E_0(N_t)/N_t$ one finds that the critical temperature in units of the string tension is infinite in the continuum limit, $T_c /\sqrt{\sigma} \to \infty$, which is the correct result. Recall from the previous discussion earlier in this section that the Nambu-Goto formula incorrectly gave $T_c /\sqrt{\sigma} \to \mathrm{constant}$.

One interesting feature of eq.\eqref{eq:aharony prediction} is that it has been derived with the assumption that $m^2 \ll \sigma$, where $m$ is the massgap and $\sigma$ the string tension. As we have seen in Section \ref{sec:threed u(1)}, in particular eq.s\eqref{eq:mass prediction}-\eqref{eq:string tension prediction}, on the lattice one expects $\frac{m^2}{\sigma} \sim (am) \beta$ up to some constants which are difficult to control at finite lattice spacing. As we have seen in Section \ref{sec:threed u(1)}, the mass $(am)$ decreases exponentially with $\beta$. Thus we expect eq.\eqref{eq:aharony prediction} to be valid for sufficiently small $(am) \beta$, but in this range numerical simulations are challenging due to an exponentially large correlation length $1/(am)$. Using the extrapolation eq.\eqref{eq:mass extrapolation} from \cite{TepAthen}, one expects the formula to be valid from $\beta \gtrsim 3.0$ and that it is unlikely to be valid for $\beta \lesssim 2.5$. For smaller $\beta$s, one would then expect corrections to eq.\eqref{eq:aharony prediction}. However, it is also interesting to note that, unlike ordinary predictions from effective string theory, in this case one has control over the microscopic values of the parameters. One manifestation of this fact is that corrections of higher order in $1/N_t$ to eq.\eqref{eq:aharony prediction} should be negligible for large enough $\beta$.

Another interesting prediction derived in \cite{Aharony:2024ctf} is that the string tension as a function of $\beta$ (i.e. eq.\eqref{eq:string tension prediction}) should receive an additive correction proportional to the mass squared $(am)^2$, i.e.
\begin{equation}
    \label{eq:aharony string tension correction}
    a^2 \sigma = \kappa_1 \frac{(am)}{\beta} - \kappa_2 (am)^2 \ , 
\end{equation}
with positive constants $\kappa_i$. The first term is just the standard prediction, eq.\eqref{eq:string tension prediction}, which we discussed in Section \ref{sec:threed u(1)}. The second term is the correction predicted in \cite{Aharony:2024ctf}. We were unable to find this correction in a reanalysis of available data \cite{TepAthen, DifferentString} at $1.9 \leq \beta \leq 2.4$, but this should be expected; in fact in this range of couplings, the parameter $(am)^2/(a^2 \sigma)$ (which is assumed to be small) is rather large, i.e. $\gtrsim 2.51$. We did not attempt to take numerical data to check this further. 

\section{Results}
\label{sec:results}

\begin{table}[bp]
\begin{tabular}{|c|c c c c c c |}
\hline
\diagbox[width=1.3cm]{$\beta$}{$N_t$} & 6 & 8 & 10 & 12 & 14 & 16\\
\hline
$1.7$
& $0.64277(45)$
& $0.91381(27)$
& $1.17388(26)$
& $1.42858(25)$
& / & / \\
$1.9$
& $0.30677(49)$
& $0.46499(26)$
& $0.61365(25)$
& $0.75642(24)$ 
& / & / \\
$2.0$
& $0.20434(48)$
& $0.32534(26)$
& $0.43885(24)$
& $0.54696(23)$
& / & /\\
$2.2$ &  /
& $0.15337(26)$
& $0.21993(24)$
& $0.28284(23)$
& $0.34354(33)$
& $0.40316(38)$\\
\hline
\end{tabular}

\vspace{0.5cm}

\begin{tabular}{|c|c c c c c|}
\hline
\diagbox[width=1.3cm]{$\beta$}{$N_t$} & 10 & 12 & 14 & 16 & 18\\
\hline
$2.4$
& $0.10568(16)$
& $0.14316(15)$
& $0.17976(34)$
& $0.21434(41)$
& $0.24943(41)$\\
\hline
\end{tabular}

\vspace{0.5cm}

\begin{tabular}{|c|c c c c c c|}
\hline
\diagbox[width=1.3cm]{$\beta$}{$N_t$} & 18 & 20 & 22 & 24 & 26 & 28\\
\hline
$3.0$
& $0.0324(17)$
& $0.04310(81)$
& $0.0501(13)$
& $0.0572(10)$
& $0.0625(12)$
& $0.0708(12)$\\
\hline
\end{tabular}

\caption{Simulation parameters $\beta$ (bare coupling) and $N_t$ (number of sites in the time direction) together with the fitted values of the ground state energy of the string $E_0$ at given $\beta$ and $N_t$, as described in the main text. A slash indicates no data was taken at that point. \label{tab:parameter values}}
\end{table}

As described in Section \ref{sec:threed u(1)}, we performed numerical simulations on lattices of size $N_t L^2$ using the snake algorithm, and obtained numerical data for the ratio of Polyakov loop correlators $Z(r+1)/Z(r)$. By experimentation, we found that $L=8N_t$ was sufficient to suppress finite-volume effects and to ensure that the volume is sufficiently large to accommodate the string tension and mass, i.e. $L\sqrt{\sigma} \gtrsim 8$ and $L m \gtrsim 8$. In order to reduce the cost of the simulations, we did not compute all the values $Z(r+1)/Z(r)$, rather we took one value every four in the range $N_t \leq r < L/2$. We checked that the systematic error due to this choice is significantly smaller than the statistical error, although it results in larger errors on the measured value of the energy. 

\begin{figure}[htbp]
\centering
\includegraphics[width=0.7\textwidth]{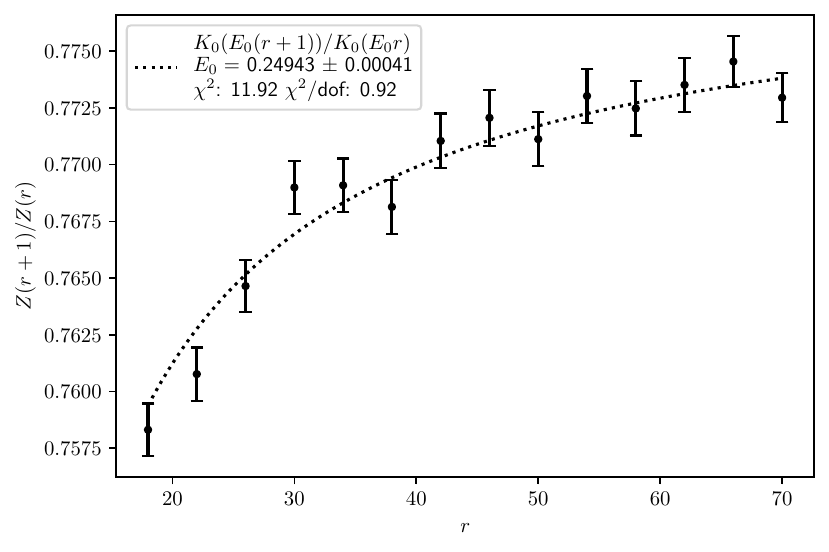}
\caption{Fitting the Monte Carlo data for the ratio of Polyakov loop correlators $Z(r+1)/Z(r)$ to the ratio of Bessel functions eq.\eqref{eq:one state ratio fit}, at $\beta=2.4$, $N_t=18$. \label{fig:bessel ratio fit}}
\end{figure}

Due to the BKT nature of the deconfinement transition in this theory \cite{Svetitsky1982}, it is difficult to clearly establish the location of the deconfinement transition \cite{Coddington:1986jk, Chernodub1, Borisenko:2008sc, Borisenko:2010qe, Borisenko:2015jea}. For our purposes, this is not however directly relevant; we are satisfied that at the range of $\beta$ and $N_t$ that we have considered, the theory is in the confined phase. Since the critical coupling $\beta_c$ grows proportionally to $N_t$, at larger $\beta$ we have to consider larger $N_t$ in order to stay in the confined phase.

We have summarized the measured values of $E_0(N_t)$ as well as the simulation parameters $\beta$ and $N_t$ in Table \ref{tab:parameter values}. The ground state energy $E_0(N_t)$ was extracted as follows. We fitted the ratio of Polyakov loop correlators to the prediction eq.\eqref{eq:luscher weisz correlator} truncated to the first term,
\begin{equation}
    \label{eq:one state ratio fit}
    \frac{Z(r+1)}{Z(r)} = \frac{K_0((r+1)E_0 )}{K_0(rE_0)} \ .
\end{equation}
For all the points except those at the largest coupling $\beta=3.0$, we found our data to be in good agreement with eq.\eqref{eq:one state ratio fit} in the whole range $N_t \leq r \leq R$. This is shown for one example in Fig.~\ref{fig:bessel ratio fit}. Only for the smallest two values of $N_t$ at $\beta=2.4$ we had to exclude a few small points near $r = N_t$, which however does not affect the value of $E_0$ within error; this is to be expected since these points are close to the deconfinement transition.

\begin{figure}[htbp]
\centering
\includegraphics[width=0.7\textwidth]{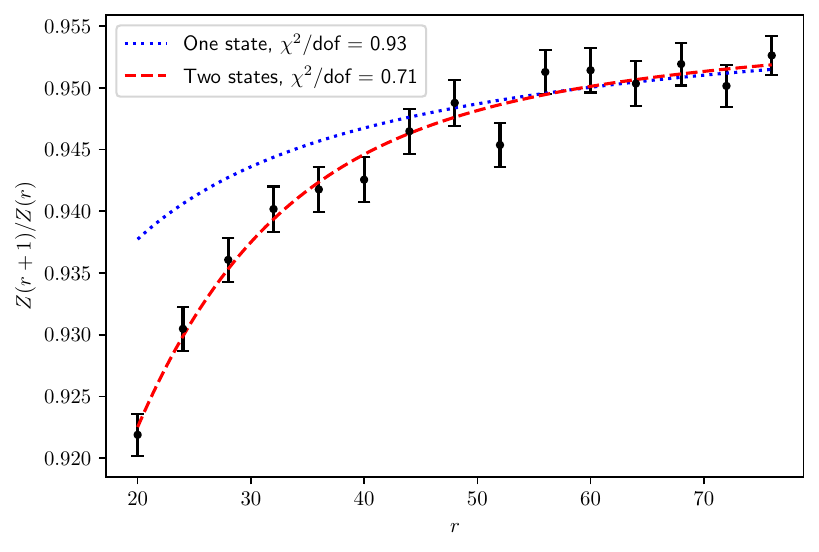}
\caption{Comparison between the one state fit with eq.\eqref{eq:one state ratio fit} (\textit{dotted blue}) and the two state fit with eq.\eqref{eq:two state ratio fit} (\textit{dashed red}) to the Monte Carlo data for the ratio of Polyakov loop correlators $Z(r+1)/Z(r)$, at $\beta=3.0$, $N_t=20$. For the two state fit, no points were excluded. For the one state fit, a comparable $\chi^2/\mathrm{dof}$ could only be obtained by excluding six points at small $r$. \label{fig:bessel ratio fit comparison}}
\end{figure}

For the points at $\beta=3.0$, agreement with eq.\eqref{eq:one state ratio fit} is found only by excluding many points at smaller $r$ (i.e. those close to $r \approx N_t$). We interpret this as being due to excited state contamination, so we have considered the fit form
\begin{equation}
    \label{eq:two state ratio fit}
    \frac{Z(r+1)}{Z(r)} = \frac{E_0 K_0((r+1) E_0)+ C E_1 K_0((r+1)E_1)}{E_0 K_0(r E_0)+ C E_1 K_0(r E_1)} \ ,
\end{equation}
which is obtained by truncating eq.\eqref{eq:luscher weisz correlator} to the first two terms. We fit the three parameters $E_0, E_1, C$ where $C$ must be positive.  In this case, we obtain an excellent fit over the whole range $N_t \leq r < L/2$ without excluding points. A comparison of the fits with one and two states is shown in Fig.~\ref{fig:bessel ratio fit comparison}. The reported values of $E_0$ in Table \ref{tab:parameter values} are those for the fit with two states, but in any case the difference with the fit with one state (excluding several smaller $r$) is at most one standard deviation. For the smaller values of $\beta$ we did not find evidence of a second state.

Having extracted the values of $E_0(N_t)$ for several $\beta$s and $N_t$s, the next question is to understand the dependence of $E_0(N_t)$ on $N_t$. At least four theoretically motivated fit forms are possible: the first one is the Nambu-Goto formula eq.\eqref{eq:E0 Nambu-Goto}; the second one is the Nambu-Goto formula with non-universal corrections eq.\eqref{eq:gamma3 gamma5 corrections}; the third one is the formula by Aharony, Barel, Sheaffer, eq.\eqref{eq:aharony prediction}; the fourth is the rigid string prediction eq.\eqref{eq:rigid string prediction}.

\begin{figure}[htbp]
\centering
\includegraphics[width=0.7\textwidth]{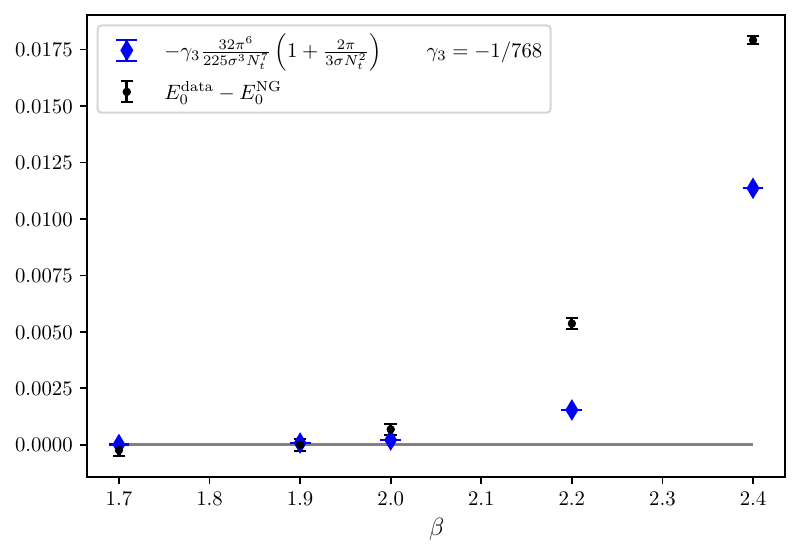}
\caption{Difference between the measured value of the string ground state energy $E_0^\mathrm{data}$ and the Nambu-Goto prediction $E_0^\mathrm{NG}$, eq.\eqref{eq:E0 Nambu-Goto}, (\textit{black circles}) at fixed $N_t=10$ and varying $\beta$, together with the maximal correction due to the $\gamma_3$ term in eq.\eqref{eq:gamma3 gamma5 corrections} up to order $1/N_t^9$ (\textit{blue diamonds}). The Nambu-Goto values are computed using the zero-temperature string tension from \cite{DifferentString}. The horizontal gray line is located at $y=0$. For smaller $\beta$, the data is in perfect agreement with Nambu-Goto. However, there is clear disagreement at the larger values of $\beta$. The correction due to $\gamma_3$ is insufficient to explain the discrepancy between the data and the Nambu-Goto prediction at $\beta=2.2, 2.4$. \label{fig:comparison fixed Nt=10}}
\end{figure}

With the exception of $\beta=3.0$, for all the other couplings we considered, the zero-temperature string tension and mass are known to high precision \cite{DifferentString, TepAthen}. Of course, one should always keep in mind that the determination of the zero-temperature string tension depends on the chosen fit form, which is also the problem we are trying to tackle in this work. However we note that the values of $\sigma$ and $m$ obtained in \cite{DifferentString} and \cite{TepAthen} are compatible with each other within error. 

Using the zero-temperature string tension, one can compare the numerical data for $E_0(N_t)$ with the Nambu-Goto (NG) prediction obtained by plugging the zero-temperature string tension into eq.\eqref{eq:E0 Nambu-Goto} without fitting. In Fig.~\ref{fig:comparison fixed Nt=10} we plot the difference $E_0^\mathrm{data}-E_0^{\mathrm{NG}}$ at fixed $N_t=10$ at varying $\beta$ in the range $1.7 \leq \beta \leq 2.4$. The qualitative features of the plot are the same for all the values of $N_t$ where we have enough data points for a comparison. As is clear from the plot, for the smaller values $\beta=1.7, 1.9$ the Nambu-Goto prediction is in perfect agreement with the numerical data. At $\beta=2.0$ we start seeing deviations from Nambu-Goto behavior; at the larger couplings $\beta=2.2, 2.4$ the numerical data is incompatible with the Nambu-Goto prediction by several standard deviations. Two features of Fig.~\ref{fig:comparison fixed Nt=10} are especially interesting. First of all, the deviation from Nambu-Goto becomes increasingly larger at larger $\beta$. The second feature is that the difference $E_0^\mathrm{data}-E_0^{\mathrm{NG}}$ is \textit{positive}, i.e. the data lies above Nambu-Goto. In other words, the correction to Nambu-Goto must have overall positive sign.

\begin{figure}[htbp]
\centering
\includegraphics[width=0.7\textwidth]{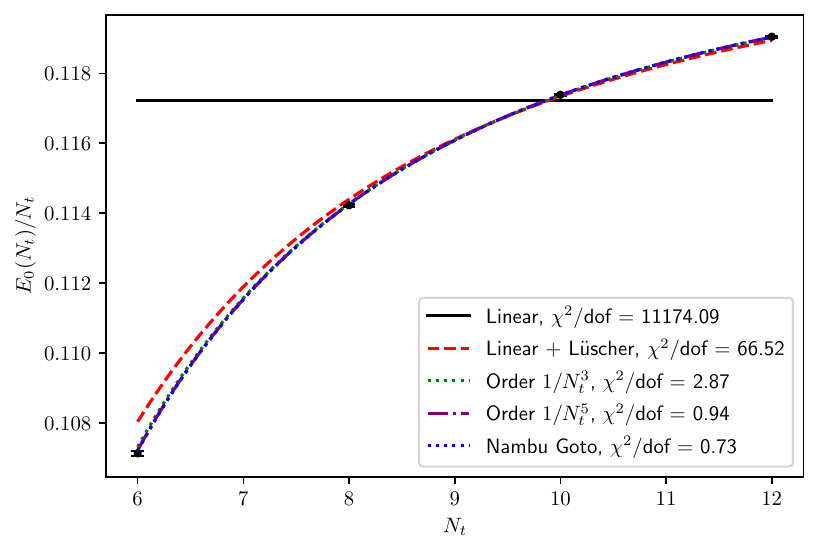}
\caption{Fit of the ground state energy $E_0(N_t)$ as a function of $N_t$ to the Nambu-Goto prediction eq.\eqref{eq:E0 Nambu-Goto} at $\beta=1.7$. The other fit forms are successive approximations of the Nambu-Goto formula from its Taylor expansion. For ease of visualization, in the $y$ axis the energies are normalized by $N_t$. As is clear from the plot and from the $\chi^2$ values, the Nambu-Goto formula provides an excellent fit to the data. Already at order $1/N_t^5$, the fit is visually indistinguishable from the full Nambu-Goto expression.   \label{fig:Nambu-Goto fit beta 1.7}}
\end{figure}

We may then ask whether this deviation can be accounted for by non-universal $\gamma_3$ and $\gamma_5$ corrections, as given in eq.\eqref{eq:gamma3 gamma5 corrections}. Because the data lies above Nambu-Goto, to have any chance of explaining the data $\gamma_3$ must be negative. Due to the inequality eq.\eqref{eq:gamma_i inequalities}, $\gamma_3 \geq -1/768$; therefore the maximal positive correction due to the $\gamma_3$ term is obtained when $\gamma_3 = -1/768$.
The maximal correction due to $\gamma_3$ up to $1/N_t^9$ (i.e. the second and third terms in eq.\eqref{eq:gamma3 gamma5 corrections}, with $\gamma_3=-1/768$) is also shown in Fig.~\ref{fig:comparison fixed Nt=10}. We see that the correction goes in the right direction but it is insufficient to account for the discrepancy between Nambu-Goto and the numerical data.
Moreover, adding the $\gamma_5$ term (i.e. order $1/N_t^{11}$ in eq.\eqref{eq:gamma3 gamma5 corrections}) does not improve the situation; due to the inequalities eq.\eqref{eq:gamma_i inequalities}, a tedious but elementary calculation shows that the maximal positive correction due to the sum of the $\gamma_3$ and $\gamma_5$ terms (up to $1/N_t^{11}$) is smaller than the one given by the $\gamma_3$ term alone (up to $1/N_t^{9}$). In fact, if $\gamma_3=-1/768$ then $\gamma_5$ is forced to be positive, thus decreasing the value of the correction.

It is also clear that the data are incompatible with the expression given by Nambu-Goto plus the rigid string correction of eq.\eqref{eq:rigid string prediction} which, as we mentioned above, is \textit{negative}, while, as we have seen, our data shows \textit{positive} deviations on top of Nambu-Goto.

\begin{figure}[htbp]
\centering
\includegraphics[width=0.7\textwidth]{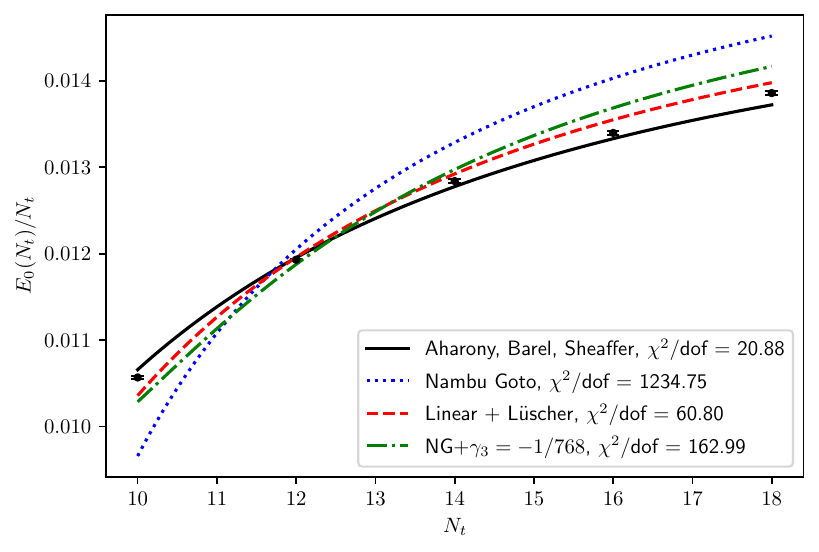}
\caption{Fit of the ground state energy $E_0$ as a function of $N_t$ at $\beta=2.4$ with eq.\eqref{eq:aharony prediction} (Aharony, Barel, Sheaffer, \textit{solid black}) where we fix $m$ to its physical value and fit $\sigma$, eq.\eqref{eq:E0 Nambu-Goto} (Nambu-Goto, \textit{dotted blue}), simply a linear term plus the L\"uscher correction (\textit{dashed red}) as well as Nambu-Goto with non-universal corrections, eq.\eqref{eq:gamma3 gamma5 corrections}, with fixed $\gamma_3=-1/768$ (\textit{dashed-dotted green}). For ease of visualization, in the $y$ axis the energies are normalized by $N_t$. It is clear that the data are not well described by any of the fit forms.} \label{fig:plot all together 2.4}
\end{figure}

The overall picture is therefore that Nambu-Goto is in excellent agreement with the numerical data for small $\beta=1.7, 1.9$, but for increasing $\beta$ it is increasingly incompatible with the data. This had also been observed in high-precision simulations at zero temperature \cite{DifferentString}. We now discuss our fits of the ground state energy $E_0(N_t)$ as a function of $N_t$ at fixed $\beta$. We separate the discussion for the three regimes low $\beta$, intermediate $\beta$ and large $\beta$.

\subsection{Low \texorpdfstring{$\beta$}{beta} regime: \texorpdfstring{$\beta=1.7~,~1.9$}{beta=1.7,1.9.}}

As shown in Fig.~\ref{fig:Nambu-Goto fit beta 1.7} at $\beta=1.7$, for the smaller values of $\beta$ the fit with the Nambu-Goto formula eq.\eqref{eq:E0 Nambu-Goto} is excellent. In the same figure, we have also fitted the same data with successive approximations obtained by Taylor expanding the Nambu-Goto formula. Already at order $1/N_t^5$, the fit are visually indistinguishable from the full Nambu-Goto expression and the reduced $\chi^2$ are excellent. The zero-temperature string tension obtained with the Nambu-Goto fit is $\sigma=0.12273(1)$, which is compatible with previous determinations \cite{TepAthen, DifferentString}. At these couplings, possible subleading corrections to Nambu-Goto are much too small to be visible; this is in agreement with the observations of the previous section, as well as with the data plotted in Fig.~\ref{fig:comparison fixed Nt=10}. As is clear from comparing the data with the linear fit in Fig.~\ref{fig:Nambu-Goto fit beta 1.7}, the string energy displays significant deviations from linear behavior.
 
\subsection{Intermediate \texorpdfstring{$\beta$}{beta} regime: \texorpdfstring{$\beta=2.0~,~2.2$}{2.4} and \texorpdfstring{$2.4$}{2.4}.}

For the intermediate values of $\beta$, none of the fit forms given by eq.s \eqref{eq:E0 Nambu-Goto}, \eqref{eq:gamma3 gamma5 corrections}, \eqref{eq:rigid string prediction} or \eqref{eq:aharony prediction} (or various combinations thereof) fit the data appropriately. 
In Fig.~\ref{fig:plot all together 2.4} we show the best fit functions for each of the fit forms superimposed on the same plot. For the prediction by Aharony, Barel, Sheaffer, eq.\eqref{eq:aharony prediction}, we fixed the mass to its physical value computed in \cite{TepAthen, DifferentString}, while for the non-universal corrections to Nambu-Goto eq.\eqref{eq:gamma3 gamma5 corrections} we fixed $\gamma_3=-1/768$ as expected from the discussion in the previous section. Thus in all cases only the string tension is left as a free fit parameter.
As is clear from the plot, and unlike what we saw at smaller $\beta$s, the Nambu-Goto prediction is incompatible with the numerical data. This we had already anticipated in the discussion surrounding  Fig.~\ref{fig:comparison fixed Nt=10}. Moreover, neither the linear plus L\"uscher formula nor the prediction by Aharony, Barel, Sheaffer have an acceptable reduced $\chi^2$. However, the latter is somewhat preferable due to its smaller reduced $\chi^2$; it gives an acceptable value of the string tension, i.e. $\sigma = 0.01529(1)$, compatible with the value in \cite{TepAthen}, i.e. $0.01541(22)$. The prediction given by the non-universal corrections to Nambu-Goto eq.\eqref{eq:gamma3 gamma5 corrections} (where we fixed $\gamma_3=-1/768$), although an improvement over plain Nambu-Goto, is also incompatible with the data and fares worse than either linear plus L\"uscher or the prediction by Aharony, Barel, Sheaffer. Leaving $\gamma_3$ as a free fit parameter results in a somewhat improved fit, but with values of $\gamma_3$ which violate the inequalities \eqref{eq:gamma_i inequalities}; this had already been anticipated at the beginning of this section. On the other hand, attempting to fit the data with the rigid string prediction \eqref{eq:rigid string prediction} (not shown) leads to very large values of the parameter $\widetilde{m}$, for which the correction essentially vanishes. That the rigid string was incompatible with our data had already been observed in the previous section. Overall, we find that our data for $\beta=2.2, 2.4$ cannot be reasonably described by any of the fit forms we considered.

\begin{figure}[htbp]
\centering
\includegraphics[width=0.7\textwidth]{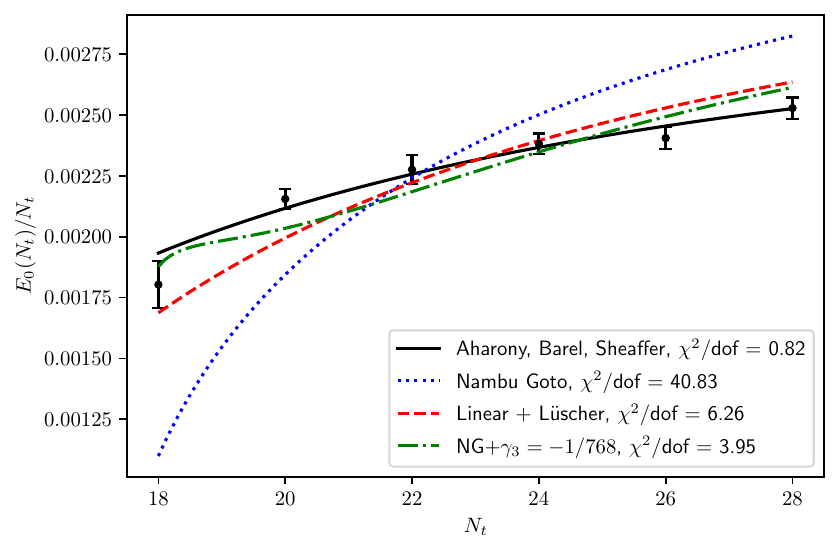}
\caption{Same as Fig.~\ref{fig:plot all together 2.4}, but at $\beta=3.0$. The value of $m$ in the prediction by Aharony, Barel, Sheaffer was fixed to the central value of the extrapolation eq.\eqref{eq:extrapolations 3.0}. The prediction by Aharony, Barel, Sheaffer fits the data very well and is clearly preferred over the other options. \label{fig:plot all together 3.0}}
\end{figure}

\subsection{Large \texorpdfstring{$\beta$}{beta} regime: \texorpdfstring{$\beta=3.0$}{beta=3.0}.}

In order to properly test the prediction eq.\eqref{eq:aharony prediction} by Aharony, Barel, Sheaffer we decided to take several data points at $\beta=3.0$ which we expect to be within the range of validity of their formula. The simulations at these values are very challenging because of the large correlation length. No results are available in the literature on the zero-temperature mass and string tension at $\beta=3.0$. The extrapolations of \cite{TepAthen}, reported in eq.s \eqref{eq:mass extrapolation}-\eqref{eq:string tension extrapolation} give the values
\begin{align}
    \label{eq:extrapolations 3.0}
    m &= 0.0382(19) \ ,  \\
    \sigma &= 0.00306(38) \ .
\end{align}
We can use these to compare with the fitted values that we measure. In Fig.~\ref{fig:plot all together 3.0} we show, as before, the various best fit lines superimposed on the same plot. As anticipated, at this large coupling the simulations are more difficult and therefore the errors are larger. However, they are still sufficiently small to allow a discrimination between the various fit forms. Unsurprisingly, Nambu-Goto is unable to describe the measured data. However, in this case it is clear both visually and by looking at the reduced $\chi^2$, that the prediction eq.\eqref{eq:aharony prediction} by Aharony, Barel, Sheaffer fits the data very well, with a visually excellent fit and small reduced $\chi^2$. Again, this is only a one parameter fit, since we have fixed the mass to its physical value given by the extrapolations \eqref{eq:extrapolations 3.0}; the fitted value of the string tension, $0.00305(2)$, is in excellent agreement with the value expected from the extrapolation eq.\eqref{eq:extrapolations 3.0}. Thus we find the prediction by Aharony, Barel, Sheaffer to be in excellent agreement with the data at $\beta=3.0$. 

\section{Conclusions}
\label{sec:conclusions}

In this work, we computed the ground state energy of the confining string in three-dimensional $\U(1)$ gauge theory and compared it with various numerical predictions. Contrary to the behavior of non-Abelian gauge theories, the string cannot be described by small corrections around the Nambu-Goto string.

Instead, we found three different regimes. For small values of the bare coupling, far from the continuum, the potential is well-described by the Nambu-Goto string. In an intermediate range of couplings ($\beta=2.2,2.4$), we compared our data with several possible numerical predictions, and found no agreement. In particular, the string cannot be described by small corrections around the Nambu-Goto string similar to those found in non-Abelian lattice gauge theories. Finally, we performed simulations at large bare coupling ($\beta=3.0$), close to the continuum, where we would expect the formula recently derived in \cite{Aharony:2024ctf} to apply and in fact we found very good agreement between our data and their formula. This scenario is consistent with the expectation that the prediction \eqref{eq:aharony prediction} should be more reliable at larger $\beta$, as discussed in Section \ref{sec:theoretical predictions}. It seems clear, a posteriori, that the indications in favor of the rigid string found in \cite{DifferentString} were due to the fact that the simulations were performed exactly in the intermediate region where the data have a mixed, cross-over behavior as well as to the presence, in the low temperature setting of \cite{DifferentString}, of a boundary term overshadowing the fine corrections to the interquark potential. 

We would like to emphasize two crucial ingredients in our calculation without which it would have been very difficult to obtain our results: first of all, working in the closed string regime (i.e. at finite temperature in the lattice gauge theory language), where the boundary term due to the Polyakov loops can be neglected; secondly, the use of the snake algorithm in the dual formulation, which allows the study of very large correlators without an exponential decrease in the signal to noise ratio.

It would be interesting to check also the prediction of \cite{Aharony:2024ctf} for the correction to the string tension, i.e. eq.\eqref{eq:aharony string tension correction}, which however would require going to larger values of $\beta$ and for the moment is beyond our computational capabilities.
It would also be interesting to extend the same analysis to other \say{two scales} theories, like the Georgi-Glashow model or the trace deformed gauge theories in the reconfined phase.

\acknowledgments

We would like to thank Ofer Aharony for useful comments. This work was partially supported by the Simons Foundation grant
994300 (Simons Collaboration on Confinement and QCD Strings) and by the Italian PRIN ``Progetti di Ricerca di Rilevante Interesse Nazionale -- Bando 2022'', prot. 2022TJFCYB. The simulations were run on CINECA computers. We acknowledge support from the SFT Scientific Initiative of the Italian Nuclear Physics Institute (INFN).

\bibliographystyle{JHEP}
\bibliography{biblio.bib}

\end{document}